# An artificial charge-density-wave conductor realized by double quantum wells

Ting-Ting Kang[1]

**Charge-density-wave (CDW) is a modulation of the conduction electron density in a conductor. Under low temperature, it can spontaneously happen in some compounds that consist of anisotropic one-dimensional crystal structures, via a strong electron-lattice interaction mechanism.[1,2] Many celebrated phenomena, e.g. non-linear transport,[3] narrow-band noise,[4] mode-locking[5] and chaos[6] under AC voltage, etc., have been reported in CDW. However, evaluating the application potential of CDW conductors has been hampered by the inconvenient shapes and sizes of CDW single crystals.[2] Although modern fabrication technology can partly resolve those troubles, (for example, cleaved film[7] and nanowire[8] NbSe$_3$ device), the imperfections induced by fabrication that corrupt measured properties are not easy to control and estimate. Here we demonstrate a convenient CDW conductor fabricated by semiconductor double quantum wells (DQW) in a field-effect transistor (FET) configuration: a modulated electron density in one QW resulted from the charged QW nearby. The electric field dependent depinning transport characteristic of CDW is clearly present. This "artificial" CDW, capable of integrating with semiconductor industry, may give fresh impetus to revive the interests in CDW.**

Very similar to electrons–lattice coupling mechanism in inorganic CDW compound, we artificially realize CDW in 2DEG (two-dimensional electron gas) via electrons–charge coupling in a so-called "charge–CDW" scheme. The configuration is shown in Fig.1(a). In a double quantum well (QW) structures, the top QW is positively charged. Owing to the strong inter-charge repulsive force, these charges will form a charged lattice (e.g., a Wigner lattice) with lattice constant $\lambda$ to minimize the total energy. Consequently, a periodic potential distribution in low QW is developed via the capacitive coupling between two QWs. Finally, in low QW, CDW is formed by condensing electrons into those periodic potential puddles.

Fig.1(b,c) shows the experiment configuration and the optical image of device layout. The free charges in electrical isolated top QW island (TQWI), whose region is the mesa area surrounded by IG (Isolation Gate) and RG(Reset Gate), are induced by charging CG(Coupler Gate)/TQWI capacitance $C_{CG-T}$. CG, IG and RG all are prepared by 20nmTi/100nmAu (Schottky contact). The induced free charge density in TQWI is

$\rho_{charge} = -V_{CG}C_{CG-T}/(S_{TQWI}e)$, where $S_{TQWI}$= 6340 μm$^2$: area of TQWI; $V_{CG}$: the voltage on CG; $e$: charge of a free electron. This relation means the polarity and quantity of $\rho_{charge}$ can be easily controlled by $V_{CG}$. The detailed charging method is some complicate and explained in Supplementary Information.

The reset channel including Reset pad (annealed AuGeNi Ohmic contact penetrating both top and low QW, same as Source/Drain pad), acts as the reset route in Fig.1(b). A grid pattern is adopted for CG. Fig.1(d) is the used AlGaAs/GaAs DQW wafers grown by molecular-beam expitaxy on an insulative GaAs substrate. It includes a 10 nm top GaAs QW and 60nm low GaAs QW separated by 100 nm Al$_{0.3}$Ga$_{0.7}$As barrier.

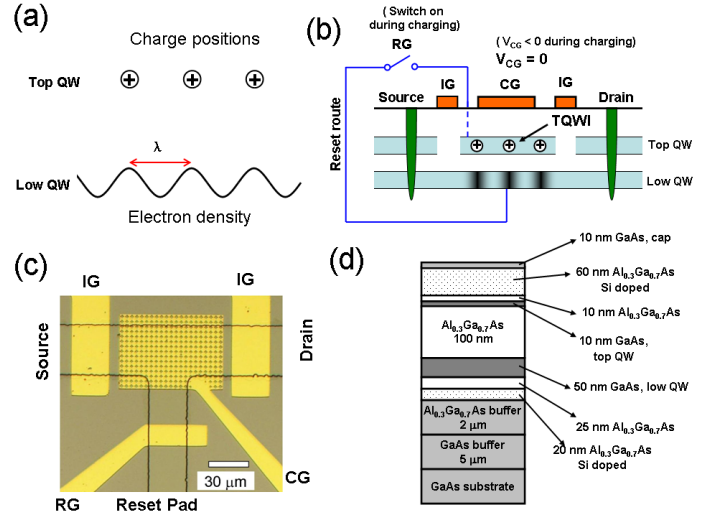

Fig. 1. (a) Schematic illustration of the CDW condensation of electrons in low QW by charges in top QW. (b) Experimental configuration showing how TQWI is charged. The dash (blue) line denotes the rest pad (An Ohmic contact connecting top and low QW).The modulated electron density in low QW from CDW formation is schematically drawn; blacker color represents higher electron density there; (c) Optical image of the device layout; (b) structures of the DQW wafers. The electron concentration and mobility for top (low) QW is 4.5×10$^{11}$ cm$^{-2}$ (3.5×10$^{11}$ cm$^{-2}$) and 7.0×10$^4$ cm$^2$V$^{-1}$s$^{-1}$ (2.1×10$^5$ cm$^2$V$^{-1}$s$^{-1}$) respectively in 4.2 K.

Simply replacing the "lattices atom" by "charges", most physics in tradition CDW crystal can be readily transferred to the charge-CDW. On the other hand, charge–CDW differs from CDW crystal in the fact: its CDW gap will not necessarily open at Fermi wave vector $k_F$, but can be arbitrarily shifted by $\rho_{charge}$

[1]National Laboratory for Infrared Physics, Shanghai Institute of Technical Physics, Chinese Academy of Sciences, Shanghai 200083, China
E-mail: ktt219@163.com .

( or $V_{CG}$ voltage). This unique property leads to the fascinating variation of CDW elastic energy, which will be heavily highlighted in the following.

Charge-CDW can be explained by the standard CDW model of Fukuyama, Lee, and Rice (FLR).[9,10] In FLR model, CDW is described as a deformable sheet (like rubber) elastically interacts with some pinning defects (impurity, surface, etc.), just like a corrugated rubber sheet moving above or pinned by a random plane of sands. The FLR Hamiltonian is:[11]

$$H = H_{elas} + H_{pin} + H_E$$
$$= \frac{1}{2}K\int dr(\nabla\phi)^2 + V_0\rho_1\sum_i \cos[Q\cdot R_i + \phi(R_i)] + \int dr \frac{\rho_{eff} E\phi}{Q} \quad (1)$$

$H_{elas}$ describes the elastic energy, where $K$ is the elastic force constant (or modulus) of the CDW; $H_{pin}$ describes the interaction with defect located at $R_i$, where $V_0$ is the defect potential, $\rho_1$ is the CDW amplitude, $Q$ is the wave vector of CDW; and $H_E$ describes the coupling of the CDW phase $\phi$ to an electric field $E$, where $\rho_{eff}$ is an effective CDW condensate density.

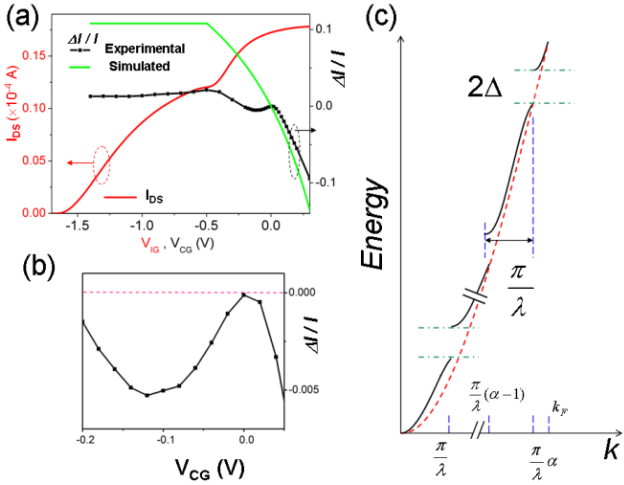

Fig.2. (a) ($\Delta I/I$)-$V_{CG}$ (black), $I_{DS}$-$V_{IG}$ plot (red),[source-drain voltage $V_{DS}$ = 8.96 mV]; simulated results by mobility-constant model (green).(b) Magnified view of ($\Delta I/I$)-$V_{CG}$ plot in -0.2V-0V. (c) The schematic dispersion relations of charge-CDW (black solid) and its origin plot before distorted by CDW gap (red dash). In (a), the derivation of simulated $\Delta I/I$ plot (green) based on mobility constant model is given in the Supplement Information.

The relative amplitude of these three terms decides the transport behavior of CDW. If $H_{elas}$ term prevails, the CDW will move freely and transport charge (Frohlich conductivity)[12] in a gapless collective manner named as phasons(a phase mode)[2], behaves like superconductors in an ideal view, but damped in reality. This case is denoted as "weak pinning". If $H_{pin}$ term prevails, CDW will enter the "strong pinning" case, a highly insulative phase due to the pinning of CDW movement by defects. Finally, $H_E$ dominates if large electric field is applied, so CDW depins from defects, carries current by CDW sliding. Thus its conduction is much enhanced compared with "weak/strong pinning" case. For this case, the CDW will behave like a trivial Ohmic conductor.

Concerning charge-CDW, to probe the transport of CDW formed in low QW, a ($\Delta I/I$)-$V_{CG}$ measurement is extensively used in this work. The meaning of $\Delta I$ and $I$ is: under a constant source-drain voltage $V_{DS}$, $I$ ($\Delta I+I$) is low QW current without (with) CDW formation respectively. So $\Delta I/I$ is the low QW current variation ratio induced by CDW. For measuring $\Delta I$, we devised a periodically four-step capacitive charging method. Very detailed explanations on this method can be found in the supplements: Fig.S1, Fig.S2 and related text. All the measurements were done in 4.2 K using liquid helium.

In Fig. 2(a), for $I_{DS}$-$V_{IG}$ plot, $V_{CG}$=0 and $V_{RG}$=0, only $V_{IG}$ is varied. Therefore, a clear double 2DEG structures is displayed and top/low QW is depleted at -0.50 V/-1.63V respectively. A source-drain voltage $V_{DS}$ = 8.96 mV is used throughout Fig.2 (a, b). We know that, when TQWI is charged, due to the capacitive coupling, electron density in low QW (below TQWI) $n_L$ will change to $n_L + \rho_{charge}$. If the electron mobility in low QW $\mu_L$ is always constant, $\Delta I$ value can be easily deduced. Thus, both the $\Delta I/I$ from experiments and mobility-constant model are given in Fig.2(a). We find these two ($\Delta I/I$)-$V_{CG}$ plots agree relatively well for $V_{CG}$>0, but badly for $V_{CG}$<0, where experimental $\Delta I/I$ is significantly reduced compared with mobility-constant model expected, indicating an obvious mobility decrease. For example, by a simple estimation, at $V_{CG}$=-0.5 V, $\mu_L$ shows a ~40% decrease (See the Supplementary Information). Especially, for -0.2 V<$V_{CG}$<0V, the overall low QW conduction even falls below the original conduction at $V_{CG}$=0V, despite its electron density is enhanced.

As discussed before, CDW forming, then pinned by impurities, is responsible for this mobility decrease. This claim will be strongly supported by the electric-field dependent transport results below. Other mobility reduction mechanisms are precluded as follows. Firstly, for positively charged TQWI (i.e. $V_{CG}$<0), electrons in low QW will be attracted and move farther from the ionized donor centers below low QW. The electron wave deformation[13] explanation expects an increase of mobility, contradicting the experimental results. So it is not the reason. Secondly, Coulomb scattering[14] is unimportant because of the large distance $d \approx 100$ nm between two QWs, so that the request $k_F d$<<1 [Fermi wave vector $k_F$=(5.9 nm)$^{-1}$ and $k_F d \approx 17$ here] is not meet.

In the framework of CDW, the absence of mobility decrease for $V_{CG}$>0 is explained by the weak electron-charge coupling and resultant small CDW gap $2\Delta$. In mean-filed approximation, $\Delta = D\exp(-1/\lambda')$ ,[1] where $\lambda'$ is the dimensionless electron-charge coupling constant, $D$ is the band-width of CDW subband. In case of $V_{CG}$>0, TQWI is filled with excessive electrons. Due to repulsive electron-electron coulomb forces, these negative charges are separated far from the higher electron density puddles in low QW [In contrast, those puddles are just below the positive charges in case of $V_{CG}<0$], resulting a small $\lambda'$, so that a much smaller CDW gap $2\Delta$. Thus the pounced electrons excited across CDW gap make the CDW behavior weak, force $V_{CG}>0$ region to behave in a trivial Ohmic way (non-CDW and mobility constant).

The interesting thing brought by charge-CDW is that CDW modulus $K$ may become controllable. In an intuitive thinking, a closer packed charge array should have very high inherent inter-charge energy, and then deforming it requires higher energy. This is reflected by a large modulus $K$ of this 2D electron sheet. Fukuyama[9] stated CDW modulus $K \propto v_F$, $v_F$ is the Fermi velocity (For example, in T=0 mean-field theory, $K = \hbar v_F /(2\pi A_0)$,[15] where $A_0$ is the unit cell cross-sectional area normal to CDW wave vector). Therefore $v_F$ is what we concern in the next.

Regarding charge-CDW dispersion relation shown in Fig.2(c), CDW formation breaks the origin band into some CDW subbands. CDW gap opens at $k = \beta\pi/\lambda$ with $\beta = 1,2,3,......\alpha$ [$\lambda = (\sqrt{|\rho_{charge}|})^{-1}$ is the CDW wavelength], not the Fermi level $k_F$ as tradition CDW. And all the filled CDW subbands contribute to CDW behavior. Using the periodicity of charge-CDW, its dispersion relations in Fig.2(c) can be folded into $(-\pi/\lambda, \pi/\lambda)$ region (the reduced Zone of CDW). Because each CDW condensate subband is filled up to $\pi/\lambda$ (namely fully filled), we reach $v_F = \hbar\frac{\pi}{\lambda}/m^*_{CDW}$ ($m^*_{CDW}$ is the effective mass of CDW condensate). If $m^*_{CDW}$ being constant is not a bad approximation, it gets $K \propto v_F \propto 1/\lambda \propto |V_{CG}|$, which means the stiffness of charge CDW can be enhanced by the charge density in TQWI or the $|V_{CG}|$ value.

From FLR model, the ratio between pinning energy $H_{pin}$ and elastic energy $H_{elas}$ is roughly $H_{pin}/H_{elas} \approx \xi = V_0\rho_1/(\hbar v_F n_i)$,[9] $n_i$ is the defect concentration. For $V_{CG} < 0$ region in Fig.2(a), pinning effect characterized by $H_{pin}$ is only important when $|V_{CG}|$ is small, i.e. $-0.12V < V_{CG} < 0$ [Fig.2(b)]. At that time, a downward evolution of transport ability, totally contrary to constant-mobility model, reflects the motion of CDW being seriously pinned. When $V_{CG} < -0.12V$, pinning effect is weakened by strong elastic energy $H_{elas}$, so that a upward evolution of $\Delta I/I$ is observed, indicating a recovery of CDW mobility against pinning defects.

The "sliding" motion[3,4] is the most characteristic behavior of CDW. Simply speaking, in an electric field dependent transport experiment, above the threshold electric field $E_T$ (or equally the threshold source-drain voltage $V_{DS,T}$), CDW will suddenly be liberated from pinning centers, and slides relative to lattice atoms (or charges in our case). This sliding triggers a collective charge transfer, producing an increased current $\Delta I$.

We have also done such electric field dependent experiment [Fig.3(a)] by varying source-drain voltage $V_{DS}$. It is found that the device conductivity (or $\Delta I/I$ ratio) is really increased in large electric field. In $V_{CG} < 0$ region, an increase of $V_{DS}$ significantly enhance $\Delta I/I$ ratio. For example, $\Delta I/I < 0$ regime with a range $-0.2V < V_{CG} < 0$ under $V_{DS} = 0.67$ mV is completely reversed to $\Delta I/I > 0$ under $V_{DS} = 65.26$ mV. The enhanced conductivity by large electric field supports CDW pinning effect as the suitable mobility reduction mechanism, while rejecting other mechanisms, as we claimed before.

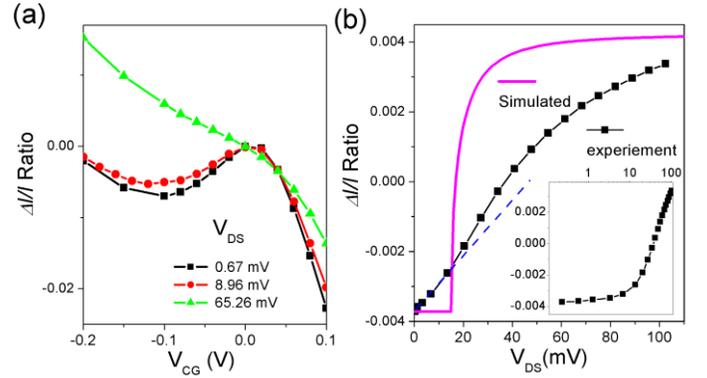

Fig. 3. (a) ($\Delta I/I$) Vs. $V_{CG}$ plot with $V_{DS}$=0.67 mV, 8.96 mV and 65.26 mV respectively; (b) With $V_{CG}$= -0.04 V, experimental ($\Delta I/I$) Vs. $V_{DS}$ plot (black) and the simulated plot by standard CDW model (pink). The straight (blue) dash line is a guide to the eye, showing the sudden increase of $\Delta I/I$ value. The inset is the experimental ($\Delta I/I$) Vs. $V_{DS}$ plot with $V_{DS}$ in log scale. The simulated plot in (b) is produced by [adapted from Eq.(5.3) in Ref.(1)]:

$$\frac{\Delta I/I}{x_L} \approx \Delta\sigma_{TQWI}/\sigma_{TQWI} = \frac{n_N + n_{CDW}(1-V_{DS,T}^2/V_{DS}^2)^{1/2}}{n_0} - 1, \quad \text{if } V_{DS} > V_{DS,T};$$

$$\frac{n_N}{n_0} - 1, \quad \text{if } V_{DS} < V_{DS,T}.$$

Where $x_L = 4$ is the length ratio between total source-drain channel and TQWI (See the supplementary information for more details), $n_0$ is the original electron density in low QW when $\rho_{charge} = 0$. The fitting values are $[(n_N/n_0)-1]/4 = -0.00372$, threshold voltage $V_{DS,T} = 15$mV, also the relation $n_{CDW} = n_0 + |\rho_{charge}| - n_N$, where $n_{CDW}$ ($n_N$) is CDW condensed (uncondensed) electron density respectively in low QW.

In Fig.3(b), keeping $\rho_{charge}$ constant ($V_{CG}$=-0.04V), $E_T$ can be roughly estimated by $\Delta I/I$ Vs. $V_{DS}$ plot. Some surprisingly, there is a linear response of $\Delta I/I$ to $V_{DS}$. A deviation (i.e. visible increase) of $\Delta I/I$ from the linear response is observed at $V_{DS}$ around 15mV, which is attributed to $E_T$. The linear background of $\Delta I/I$ indicates that charge-CDW already starts to slide under small electric field, not strictly needs $E > E_T$. We explain this feature as a result from non-uniformity of CDW. This can possibly happen if phase-phase correlation length $l_\phi$ is small and the overall CDW is actually a collection of many nearly independent phase-coherent CDW domains.

In FLR model, minimizing the sum of 2D CDW elastic energy and pinning-energy per unit are of size $l_\phi^2$ yields[11]

$$l_\phi = (\frac{\pi^2 K}{V_0 \rho_1 \sqrt{n_i}})\sqrt{t} \quad (2)$$

where $t$ is 2D CDW thickness. We already have $K \propto 1/\lambda \propto |V_{CG}|$ so that $l_\phi \propto |V_{CG}|$. It means that we will have a less number but larger sized CDW domains under large $|V_{CG}|$. The ideal case is that the device consists of only one CDW domain, whose $E_T$ will just be $E_T$ of the device. In this case, same as what the standard CDW theory predicts [the simulated (in pink) plot in Fig.3(b)], a sharp increase of $\Delta I/I$ when $V_{DS} > V_{DS,T}$ (the typical threshold behavior) is expected. While the non-uniformity brought by many small sized CDW domains with different properties will make the threshold behavior less visible. Those situations are just what we observe in Fig.4. For $V_{CG}$= -0.5 V, the $\Delta I/I$ increase is rather sharp near $V_{DS}$ ~20mV. While if $|V_{CG}|$ decrease to its 1%, i.e. $V_{CG}$ =-0.005 V, only the linear background is visible in all $V_{DS}$ range.

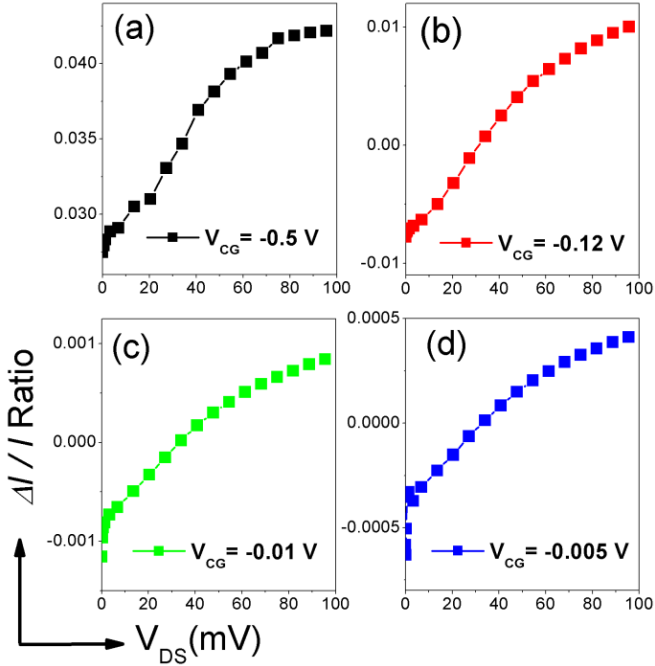

Fig. 4, Experimental ($\Delta I/I$) Vs. $V_{DS}$ plot under different $V_{CG}$ values. $V_{CG}$ values have been indicated inset. The threshold behavior of $\Delta I/I$ near $V_{DS}$~ 20 mV becomes sharper if $|V_{CG}|$ increases more and more large.

After deriving $l_\phi$, threshold electric field $E_T$ is determined from the electrical energy needed to overcome the total energy per unit area[11]

$$E_T = \frac{Q(V_0 \rho_1 \sqrt{n_i})^2}{2\rho_{eff}\phi_T(\pi^2 K)} \frac{1}{t} \quad (3)$$

where $\phi_T$ is the threshold depinning angle. We omit those uninterested term, so $E_T \propto Q/K$. Using the unique property of charge-CDW $K \propto 1/\lambda = 2\pi/Q$, we get $E_T \propto Q/(2\pi/Q) = $ constant, namely $E_T$ is unchanged against varying charge density in top QW. This is an important claim. Concerning Fig.4, we find that, in the accuracy limit of our measurements, this claim is roughly unchallenged. And threshold behavior appears always near $V_{DS}$ ~ 20 mV.

At present, it seems two-dimensional FLR model can satisfactorily explain the experimental results of charge-CDW, supporting the existence of CDW in this charged double QW system. So, in the future, it may be possible to explore the potential of CDW devices by just using the convenient semiconductor structures.

# An artificial charge-density-wave conductor realized by double quantum wells- SUPPLEMENTARY INFORMATION

Ting-Ting Kang

## (1) Charging and ($\Delta I/I$)-$V_{CG}$ measurement method

In the following, for simplicity, we use the results with $V_{CG}$=-0.04V to illustrate our measurement method. The readers can easily think out the cases with $V_{CG}$>0.

In this scheme, charging of TQWI is realized by intentionally applying some bias to the CG gate above it. For IG, a voltage $V_{IG}= V_1$ (here $V_1$= -0.7 V), whose amplitude $V_1$ can only deplete the top QW below IG, is applied. For RG, a voltage $V_{RG} = V_1+V_{pulse}$ is applied. $V_{pulse}$ is a pulse wave with frequency $2f$, pulse amplitude is - $V_1$ (See Fig. S1, middle). In the pulse duty time, $V_{RG} = 0$ V, TQWI is connect to low QW via reset route/pad. In other time, $V_{RG} = V_1$ and TQWI is always isolated. $V_{RG}$ will act as a switch on the charging/discharging of TQWI, as schematically shown in Fig.1 (b).

A square wave voltage $V_{CG}$, alternating in 0 V and $V_2$ ($V_2 =$ -0.04 V here. The $V_2$ values are the $V_{CG}$ values given in $\Delta I/I$ measurements), with frequency $f$ is applied to CG (See Fig. S1, top). $V_{CG}$ will provide the voltage for charging TQWI through the CG/TQWI capacitance.

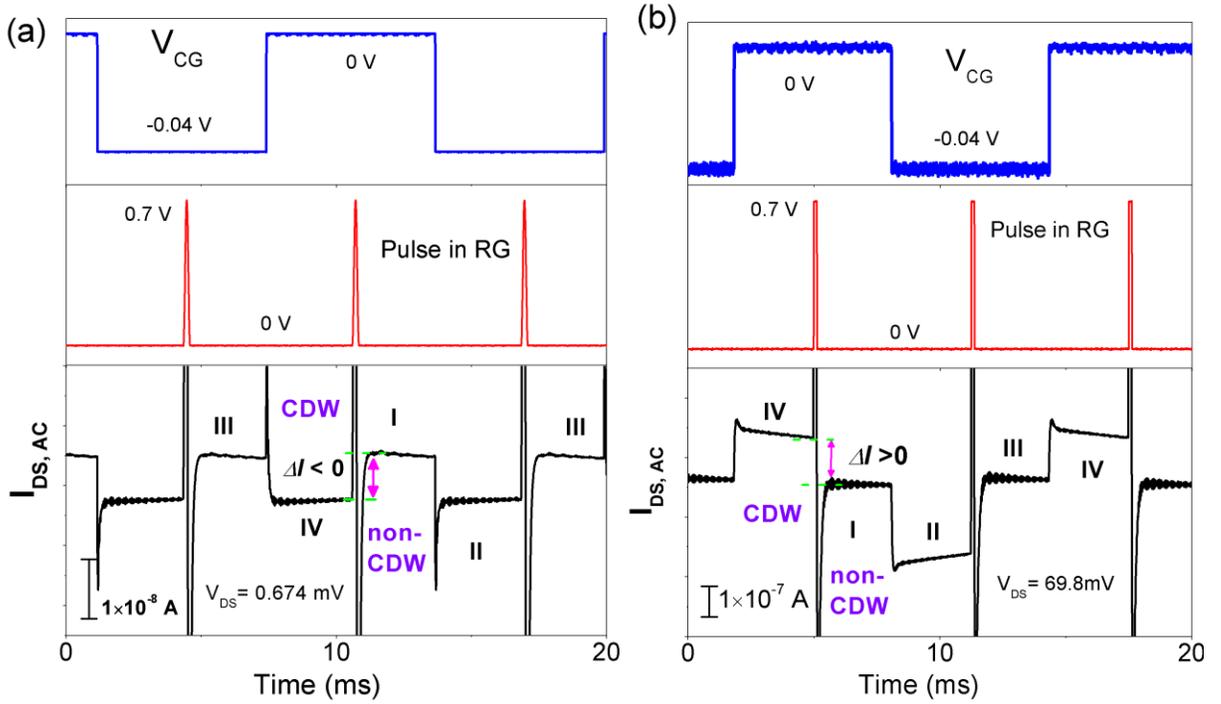

Fig. S1, One example: time traces of source-drain current of low QW in the "capacitive charging" measurement. The source-drain voltage $V_{DS}$ is (a) 0.674 mV; (b) 69.8 mV. The voltage on Coupler Gate $V_{CG}$ is -0.04V .The current change between step IV and I are figured out. The time evolution of the D-S current is monitored by observing the AC component of Source-Drain current using a four-channel digital oscilloscope (Tektronix, Inc.). The square wave in coupler gate (frequency $f$, $f$ = 80 Hz here) and pulse in reset gate (frequency $2f$, $2f$ =160 Hz here.) are both provided by a multifunction generator (NF Corporation, WAVE FACTORY).

Fig.S2 illustrates how one charging/discharging cycle is going on. Previous to the charging/discharging cycle, a negative voltage $V_1$ is applied to both IG and RG, so that we get an isolate TQWI. The charging/discharging cycle can be divided into 4 steps. The electron density for low QW in step $i$ is denoted as $N_i$. In each step, TQWI is always isolated and TQWI is only connected to low QW in the pulse between Step IV and I, Step II and III. These steps are explained below. (I), $V_{CG} = 0$ V, TQWI is neutral; (II), $V_{CG} = V_2 < 0$. Since TQWI is floating and electrons cannot flow in/out, TQWI is still neutral. This neutral TQWI means it will not screen the electric field from CG and the negative electric field from CG will pass through top QW, reaching low QW and reducing low QW electron density. Therefore $N_{II} < N_I$; (III), $V_{CG} = V_2 < 0$. Between Step II and III, a short positive voltage pulse comes into RG, TQWI is connected to low QW during the pulse and excessive electrons flow out to low QW, leaving TQWI positively charged and screening the negative electric field from CG. As a result, for Step III, only a small negative electric field from CG reaches low QW. Therefore $N_{III} > N_{II}$; (IV), $V_{CG} = 0$V, TQWI is positively charged. Due to capacitive coupling between TQWI and low QW, electron accumulates in low QW. Therefore $N_{IV} > N_{III}$. After (IV), due to a short positive voltage pulse in RG, Reset channel turn on shortly, letting electrons flowing into TQWI during pulse and neutralize TQWI. As a result, electron accumulated in low QW disappears. The situation return to (I), starting a next cycle. And we have $N_{IV} > N_I$. From the discussion above, we have $N_{IV} > N_I \approx N_{III} > N_{II}$, which also is schematic illustrated in Fig.S2.

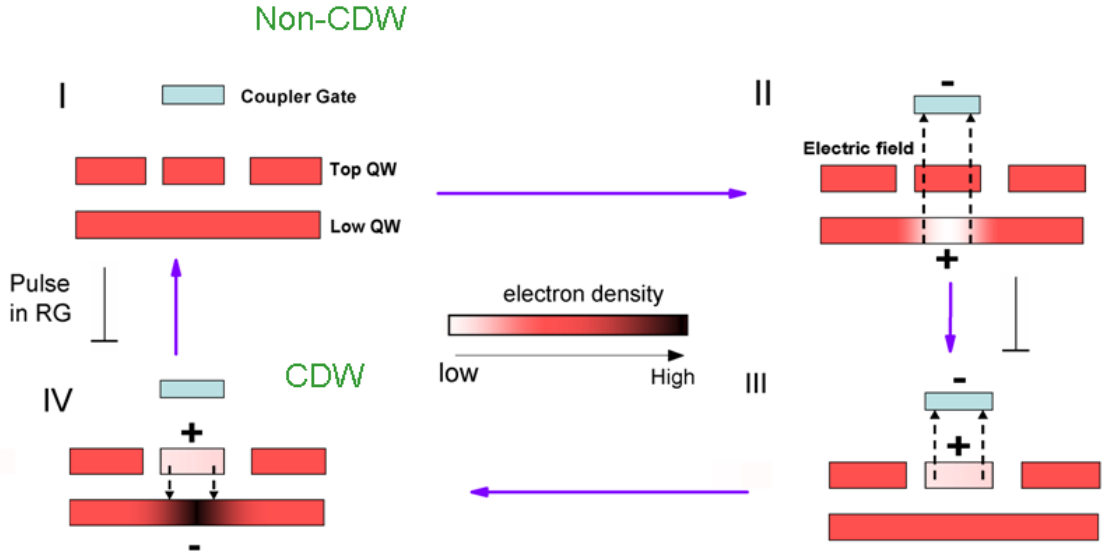

Fig. S2, Schematic illustration of one cycle of the measurement. A color bar indicating electron density is given in the center. The dash arrows denote electric field.

From the discussion above, we have $N_{IV} > N_I \approx N_{III} > N_{II}$, which also is schematic illustrated in Fig.S2. If the mobility in low QW is assumed to be constant ("mobility-constant model"), we will have $\sigma_{IV} > \sigma_I \approx \sigma_{III} > \sigma_{II}$. This is just what we observe in $V_{DS} = 69.8$ mV [Fig.S1 (b)]- the trivial case. However, for $V_{DS} = 0.674$ mV in Fig.S1 (a), $\sigma_{IV} > \sigma_I$ is reversed to $\sigma_{IV} < \sigma_I$. (This is figured out in Fig. S1.) Since $N_{IV} > N_I$ is already known, then we reach $\mu_{IV} < \mu_I$, i.e. for low QW, its electron mobility is reduced in Step IV compared with Step I, even its density is enhanced.

Finally, we give the way how ΔI and I value are deduced. As depicted in Fig. S1(low part), "ΔI" value is decided by: ΔI $= I_{DS,AC}(IV) - I_{DS,AC}(I)$; while "I" value is the DC component of source-drain current.

## (2) Derivation of the simulated *ΔI/I* plot (green) in Fig.2 (a) based on mobility constant model;

The CG/TQWI capacitance reads:
$$C_{CG-T} = S_{CG}\varepsilon_{GaAs}\varepsilon_0 / d_{CG-T}$$
where $S_I = 3200$ μm$^2$: area of CG above TQWI; $d_{CG-T} = 80$nm: distance between CG and top 2DEG; $\varepsilon_{GaAs} = 12.9$: GaAs dielectric constant; $\varepsilon_0 = 8.854 \times 10^{-12}$ F·m-1: vacuum permittivity.

The induced free charge density in TQWI is $\rho_{charge} = -(C_{CG-T}V_{CG})/(S_I e)$ [$S_I = 6340$ μm$^2$: area of TQWI; $V_{CG}$: the voltage on CG; e: charge of a free electron].

After the charging of TQWI, the electron density in low QW below TQWI $N_{LI}$ will change at a amount of $|\rho_{charge}|$ to $N_{LI} + \rho_{charge}$, due to the charged TQWI just above it. If we assume, low QW (the region below TQWI) electron mobility $\mu_{LI}$ is a constant, the low 2DEG conductivity $\sigma_{LI} = N_{LI} e \mu_{LI}$ under TQWI will change to $\sigma_{LI} = (N_{LI} - \rho_{charge}) e \mu_{LI}$.

Consider the device configuration in Fig. S3, with

$$\frac{\sigma_{LI}}{\sigma_{not-LI}} = \frac{L_{LI}^{-1}}{L_{not-LI}^{-1}} = \frac{L_{not-LI}}{L_{LI}} = \frac{L - L_{LI}}{L_{LI}} = x_L - 1 \quad \text{(Eq. S1)}$$

$$x_{\sigma,LI} = \Delta \sigma_{LI} / \sigma_{LI} \quad \text{(Eq. S2)}$$

where $x_L = L/L_{LI} = 400\mu m/100\mu m = 4$; $\sigma_{not-LI}$: Low QW conductivity of Source-drain channel except the region below TQWI; $\sigma_{LI}$ ($\sigma_{LI} + \Delta \sigma_{LI}$): Low QW conductivity of Source-drain channel under TQWI without (with) free charges in TQWI;
We have

$$\frac{I + \Delta I}{I} = \frac{\sigma'}{\sigma} = \frac{[\sigma_{not-LI}^{-1} + (\sigma_{LI} + \Delta \sigma_{LI})^{-1}]^{-1}}{[\sigma_{not-LI}^{-1} + (\sigma_{LI})^{-1}]^{-1}} = \frac{\sigma_{not-LI}^{-1} + (\sigma_{LI})^{-1}}{\sigma_{not-LI}^{-1} + (\sigma_{LI} + \Delta \sigma_{LI})^{-1}}$$

$$= \frac{\dfrac{\sigma_{LI}}{\sigma_{not-LI}} + 1}{\dfrac{\sigma_{LI}}{\sigma_{not-LI}} + \dfrac{\sigma_{LI}}{\sigma_{LI} + \Delta \sigma_{LI}}} = \frac{x_L}{x_L - 1 + \dfrac{1}{1 + x_{\sigma,LI}}}$$

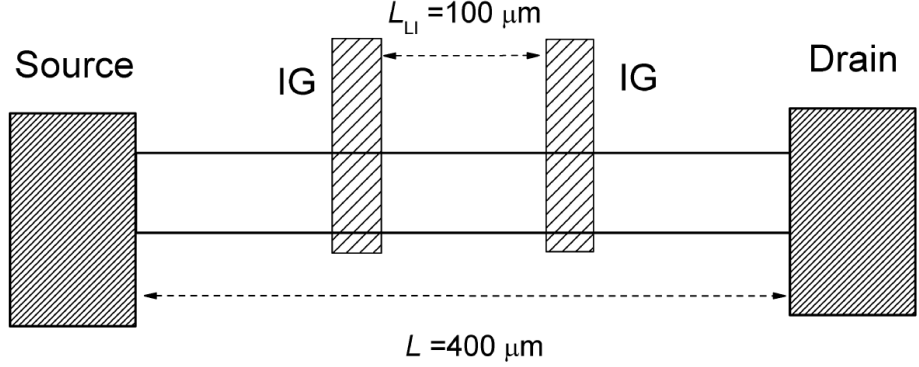

Fig. S3, Configuration of the device for deriving the simulated $\Delta I/I$ plot in Fig.2 (a)

So
$$\frac{\Delta I}{I} = \frac{I + \Delta I}{I} - 1 = \frac{x_L}{x_L - 1 + \dfrac{1}{1 + x_{\sigma,LI}}} - 1 \quad \text{(Eq. S3)}$$

When $x_{\sigma,LI} \ll 1$, a simple form is reached:
$$\Delta I / I \approx x_{\sigma,LI} / x_L = x_{\sigma,LI} / 4. \quad \text{(Eq. S4)}$$

From (Eq.S4), it is known that if $\Delta I / I$ varies not so significantly, the conductivity $\sigma_{LI}$ varying ratio can directly deduced by multiplying a length ratio 4 to $\Delta I / I$.

In Fig. 2(a,b) of the main text, at $V_{CG}$=-0.5 V, experimental $\Delta I / I$ is smaller than mobility-constant model at a rate of 10%. From (Eq.S4), we get $x_{\sigma,LI} \approx 40\%$. It means for $V_{CG}$=-0.5 V deceases at a rate of 40% compared with the case $V_{CG}$=0 V, even its electron density is larger than the case of $V_{CG}$=0 V. So it is direct to conclude that the $\mu_{LI}$-the electron mobility for low QW under TQWI, is reduced at a rate >40%.